# IMPROVED PROPAGATION MODELS FOR LTE PATH LOSS PREDICTION IN URBAN & SUBURBAN GHANA

James D. Gadze, Kwame A. Agyekum, Stephen J. Nuagah and E.A. Affum

Department of Telecommunication Engineering,
Kwame Nkrumah University of Science and Technology, Ghana

## ABSTRACT

*To maximize the benefits of LTE cellular networks, careful and proper planning is needed. This requires the use of accurate propagation models to quantify the path loss required for base station deployment. Deployed LTE networks in Ghana can barely meet the desired 100Mbps throughput leading to customer dissatisfaction. Network operators rely on transmission planning tools designed for generalized environments that come with already embedded propagation models suited to other environments. A challenge therefore to Ghanaian transmission Network planners will be choosing an accurate and precise propagation model that best suits the Ghanaian environment. Given this, extensive LTE path loss measurements at 800MHz and 2600MHz were taken in selected urban and suburban environments in Ghana and compared with 6 commonly used propagation models. Improved versions of the Ericson, SUI, and ECC-33 developed in this study predict more precisely the path loss in Ghanaian environments compared with commonly used propagation models.*

## KEYWORDS

*Propagation Models, Path loss Exponent, Root Mean Square Error, Signal Reference Received Power(RSRP).*

## 1. INTRODUCTION

Cisco's visual networking index, 2017-2022 predicts that the IP traffic recorded annually around the globe is estimated at 4.8 Zb by 2022. This translates to a threefold increase over the next five years [1]. Mobile data subscription in Ghana as of July 2018 stood at twenty-nine million, one hundred and eighty-one thousand, eight hundred and sixty-three (29,181,863) [2]. For a country with an estimated total population of thirty million [3], it shows the high demand for data and broadband services.

With this growing demand for bandwidth in mobile communication as user numbers keep increasing significantly, mobile networks have evolved from 1G - 4G to meet the demand over the years. In Ghana, Blu telecom, Busy internet, Surfline, MTN, and recently Vodafone have commercially deployed 4G LTE networks for higher throughputs and improved user experience. However, this hasn't been completely achieved since the expected throughput of 100Mbps is barely realized leading to dissatisfaction among customers. This has resulted in a lot of complaints and sanctioning from the National communications authority [4].

Transmitted signals from a base station suffer severe attenuation as they propagate through space leading to degradation in signal strength and quality [5]. This severe attenuation is introduced due to reflection, diffraction, and scattering of the signal as it impinges on obstacles. For subscribers of a network who have varying mobility, it is imperative to design a mobile network so that they have robust signal levels at all locations. To achieve this, Conditions for radio propagation need to be well understood and predicted as accurately as possible. propagation models are instrumental in wireless network planning as they support interference estimates, frequency assignments, and cell coverage assessment and other parameters[6].

Empirical propagation models that are mostly used are however environment specific and are developed based on a specific propagation environment of interest [7]. Any little deviation in characterizing the propagation environment under investigation affects the efficiency of propagation models designed from the area[8],[9]. Therefore, the use of propagation models in settings other than those intended to be used might lead to inaccurate prediction which affects system performance[10],[11]. To investigate these claims, the approach adopted in this project is to take Signal Reference Received Power (RSRP) values from deployed cell sites and compare with predictions from 4 propagation models at 800MHz and 6 models at 2600MHz. This approach will help us develop modified and improved versions of already existing propagation models suited for the Ghanaian environment.

The rest of the paper is organized as follows: Section 2 reviews some relevant work in this field. Section 3 presents the measurement procedure at investigated environments and empirical propagation models under consideration. Results are presented in Section 4 and Section 5 concludes the study.

## 2. RELATED WORKS

Considering the increased demand placed on mobile communication, higher throughputs and seamless connectivity, designing LTE networks in compliance with the performance metrics it promises is crucial. Numerous studies have gone into finding propagation models that predict accurately the path loss in the USA, Europe, Africa, and Asia to improve network performance for both voice and data communication.

How best current propagation models will perform when used in wireless environments other than those originally intended for frequently deviate from the ideal [6]. Numerous studies around the globe, however, show that many industry-standard path loss models perform effectively when adjusted to measured data from these areas [12].In [12], Path loss measured data at 3.5GHz in Cambridge was compared with the predictions of three empirical propagation models. Results indicated that the SUI and COST-231 models over-estimated path loss in this environment. The closest fit to the measurement data was the ECC-33 model. It was therefore recommended for use in urban environments.

The least-square method was used in [13] to optimize the Hata empirical path loss model for accurate prediction suited to a suburban area in Malaysia. Outdoor measurements were taken in Cyberjaya, Malaysia at a frequency range of 400MHz to 1800 MHz. Measurements were then compared with the existing models from which the Hata model showed the best fit. The optimized Hata model was used and validated in the Putrajaya region to detect the relative error to evaluate its efficiency. Smaller mean relative error was recorded hence showing that the optimization was done successfully.

Propagation models are presented in [14]  for LTE Advanced Networks.  Path loss for varying environments ( rural, suburban and dense urban)were computed using the following propagation models, COST-231 Walfisch–Ikegami model, SUI, ECC-33, Okumura extended Model and COST-231 Hata Model using MATLAB. Three frequencies between 2.3GHz and 3.5 GHz were considered in this work. Results presented indicated the COST-231 Hata model agreed better, giving the least path loss in all the environments compared with the other models. This work, however, did not compare the prediction of empirical models with measured data but only based on the model with the least path loss. The conclusion made favoring the cost 231 Hata model by simulation as agreeing best in all environments might be misleading.

Extensive measurements in [10] taken in Lagos at a frequency of 3.4GHz made a comparison with 6 standard propagation models. It was concluded that the COST 231-Hata and Ericson models showed the best performance in urban and suburban areas. Recent works in [16] also compared the efficiencies of empirical, heuristic, and geospatial methods used for signal path loss predictions using data collected in urban Nigerian cities to develop path loss models. The developed models and empirical models were compared with field measured data. All models gave acceptable RMSE values excluding the ECC-33 and Egli models. Empirical models were the simplest and most commonly applied of the three techniques submitted. Their work, therefore, emphasized the further improvement of empirical models for optimum prediction. A hybrid of heuristic and empirical models for prediction was recommended to decrease the errors associated with empirical models.

Works have also gone into comparing path loss of urban and suburban areas and to ascertain if a particular propagation model can be used for both settings.[17] showed that propagation models in urban areas experience higher losses compared with suburban areas. For all environments, no single model could be proposed.

On the background that deployed WiMAX networks, failed to meet the optimum service quality requirements for delivering continuous wireless connectivity requests in the sub-Sahara region needed for emerging mobile applications, [18] investigated the throughput performance of a deployed 4G LTE Site to ascertain if LTE meets the bandwidth demand needed for data-centric broadband applications. Field data from a deployed 4G LTE BS in Ghana operating at 2600 MHz recorded a maximum throughput of 29.9 Mbps per sector. A maximum throughput of 62.318 Mbps was recorded at the downlink for customers within 2.5 km of the cell range from the BS. It was concluded that 4G LTE can meet the ever-increasing demand of Ghanaians for broadband. This conclusion was made after comparing these throughputs with the desired throughput required to sustain datacentric broadband applications.

Works in the Ghanaian environment focusing on WiMAX networks in the 2500-2530 MHz band was presented in [11]. The measurement from a deployed WiMAX site around the university of Ghana, Accra was compared with the prediction of four empirical models. The extended COST-231 model was selected as the model that best fits the measured data because it recorded the least RMSE and a higher correlation coefficient. This model was recommended therefore for efficient radio network planning in Ghana and the sub-region at large. It was also concluded that no particular propagation model can be used to forecast coherent outcomes for all propagation settings. The reason for this was the variations in weather and geography. Recommendations were made to consider varying terrain parameters.

Intensive measurements in separate environments must be conducted to parameterize a model. The parameters of the channel model are then adjusted to suit the measurement outcomes [19]. It is imperative therefore from the works reviewed to evaluate the performance of industry-standard propagation models proposed for 4G LTE networks by considering different Ghanaian environments. With several path loss models performing differently in different environments, it is, therefore, essential to determine which of the most frequently used models is best suited for 4G LTE networks in Ghana. Further improving the suited model for more accurate prediction pertinent to the Ghanaian and Sub-Saharan environment will facilitate effective deployment of LTE networks by operators, meeting the promises the Standard came with. This, in the end, will afford subscribers the chance to enjoy seamless connectivity leading to customer satisfaction and loyalty.

## 3. MEASUREMENTS

### 3.1 Procedure

Received signal reference power (RSRP) values in dBm were taken at 10 Base stations in seven selected areas in Ghana with varying environmental conditions. A drive test was conducted using phones connected via the USB port to a computer with LTE software (Genex probe) installed on it. Genex probe serves as a data collection software interface. A GPS was attached for location finding and tracking distance covered. The frequency was set to 800MHz for the first test case at five base stations and 2600MHz for the second test case for the other five base stations. At the various sectors of each LTE site in these environments, RSRP values at a varying distance starting from a reference distance (do) of 50m to 500m with 50m intervals were recorded. The Transmit - Receiver distance was limited to 500m to reduce the impact of interference from neighboring cells and also to cater for obstructions in the way of the drive. A receiver antenna height of 1.5m was maintained throughout the measurement campaign. Measured data is sent via the phones to the computing device which stores the data as recorded log files. These recorded log files are then interpreted and analyzed. Field measurements were taken between February and May. The RSRP in dBm was taken along the LOS and NLOS of the fixed base stations with heights ranging between 16m and 35m. The laptop having GENEX software installed on it, the phone and the GPS were set up in the drive test vehicle as shown in figure 1

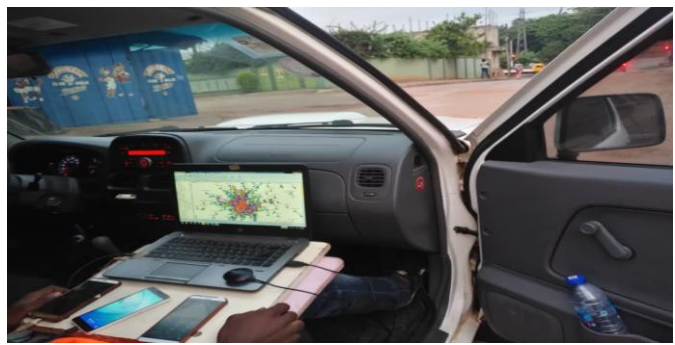

Figure 1 Measurement set up

### 3.2 Description of Environments

Drive test measurements were taken in the following environments in Ghana.

1. Adum: This is an urban area located in the central hub of Kumasi, in the Ashanti Region of Ghana with coordinates 6.6919°N,1.6287°W. It is highly populated and Characterized by a lot of business activity. Present in Adum are a lot of high-rise buildings.

2. Techiman: This is an urban area that serves as the capital of the newly created Bono East Region of Ghana with coordinates 7.5909° N, 1.9344° W. It is characterized by quite many high-rise buildings and a lot of farming and business activities.

3. Agogo: This is a Suburban area in the Asante Akyim North Municipal District of the Ashanti Region of Ghana with coordinates 6.7991° N, 1.0850° W. Agogo is approximately 80 kilometers east of Kumasi, with moderate population and buildings. Buildings are mostly not high rise and are a little isolated from each other. The terrain is relatively flat.

4.  Afrancho: It is a Populated suburban Community in the Bosomtwe District of the Ashanti region of Ghana with coordinates 6° 33' 0" N,1° 38' 0"W. It is characterized by relatively hilly terrain with the presence of valleys.
5.  New Dorma: Suburban area in the Brong Region of Ghana. It is Characterised by a mixture of flat and hilly terrains covered with a lot of vegetation. It lies on coordinates 7° 16' 39" N,2° 52' 42"W.
6.  Berekum: This is a Municipal located in the Bono Region of Ghana. It lies on coordinates 7° 27'N,2° 35'W.
7.  Sunyani: This is an Urban populated city serving as the capital of the Bono Region of Ghana. Sunyani is surrounded by the forested Southern Ashanti uplands. It lies on coordinates 7° 20'N,2° 20'W.

**Modeling Parameters**

Parameters used in generating the path loss for the different propagation models are given in Table 1.

Table 1 Modeling parameters

| parameters | | values |
|---|---|---|
| Operating frequency | | 800MHz &2600MHz |
| Transmit power | | 46dBm |
| Transmitter Antenna Height | Techiman | 35m |
| | Adum | 24m |
| | Agogo | 25m |
| | Afrancho | 32m |
| | New Dorma | 32m |
| | Berekum | 32m |
| | Sunyani | 25m |
| Shadowing factor | urban | 10.6 dB |
| | suburban | 8.2 dB |
| Distance | | 50 m − 500 m |
| Reference distance ($d_O$) | | 50m |
| Receiver antenna height | | 1.5m |

### 3.3 Propagation Models

The following propagation models were considered in this work.

I. Free Space Path Loss Model
II. Hata Model
III. COST- 231Model
IV. ECC-33 Model
V. Stanford University Interim (SUI) Model
VI. Ericson Model

#### 3.3.1  Free Space Path Loss

Path loss estimation by this model is as given in equation (1)

$$Pl = 32.44 + 20\log(d) + 20\log(f) \tag{1}$$

$f$ =frequency in MHz , $d$ =distance in Km

### 3.3.2 HATA Model

Path loss for the Hata model as given in [5] and [21] is given in (2).

$$Pl_{urban}(dB) = 69.55 + 26.16\log(f) - 13.82\log(h_t) - \partial h_r$$
$$+ [44.9 - 6.55\log(h_t)]\log(d) \tag{2}$$

in suburban areas path loss is computed as in (3)

$$Pl_{suburban}(db) = Pl_{urban} - 2[\log(\frac{f}{28})]^2 - 5.4 \tag{3}$$

$f$ =frequency in MHz, d=distance in Km, $h_r$ =mobile antenna height in meters  and  $h_t$ = base station antenna height in meters

In small and medium cities,

$$\partial h_r = (1.1\log(f) - 0.7)h_r - (1.56\log(f) - 0.8) \tag{4}$$

For large cities,

$$\partial h_r = 3.2[\log(11.75h_r)]^2 - 4.97 \text{ for f} > 300\text{MHz} \tag{5}$$

$$\alpha h_r = 8.29\left[\log\left(1.54h_r\right)\right]^2 - 1.1 \text{ for f} \leq 300\text{MHz} \tag{5}$$

### 3.3.3 COST-231 Model

The path loss  equation for this model expressed  in dB as given in [24] is shown below

PL(dB) = 46.3 + 33.9 log(f) -13.82 log($h_{te}$) -a($h_a$) + (44.9-6.55 log($h_{te}$)) log(d) + $C_m$
(6)

where $f$ is the frequency specified in $MHz$  $d$ , is the distance between the base station and mobile antennas given in km, $h_{te}$ is the base station antenna height above ground level in meters.$h_{re}$ is the mobile antenna height in meters, $C_m$ is defined as 0 dB for suburban or open environments and 3 dB for urban environments.

$a(h_a)$ is defined for large areas as

$$\alpha(h_a) \, dB = 8.29(\log(1.54h_{re}))^2 - 1.1 \, for \, f \leq 300MHz \tag{7}$$

$$\alpha(h_a) \, dB = 3.2(\log(11.75h_{re}))^2 \text{-}4.97 \text{ for f} > 300\text{MHz} \tag{8}$$

In medium or small cities,

$$\partial(h_a)dB = (1.1\log(f) - 0.7)h_r - (1.56\log(f) - 0.8) \tag{9}$$

### 3.3.4 ECC-33 Path Loss Model

The path loss equation for this  model is given in [15]

$$PL = A_{fs} + A_{bm} - G_b - G_r \tag{10}$$

Where $A_{fs}$ is the free space path loss, $A_{bm}$ is the basic median path loss $G_b$, is the transmitter antenna height gain factor and $G_r$ is the receiver antenna height gain factor.

Each of these parameters is expressed fully as;

$$Afs = 92.4 + 20log_{10}(d) + 20log_{10}(f) \tag{11}$$

$$A_{bm} = 20.41 + 9.83log_{10}(d) + 7.894log_{10}(f) + 9.56[log_{10}(f)]^2 \tag{12}$$

$$G_b = log_{10}(hb/200)\left\{13.958 + 5.8[log_{10}(d)]^2\right\} \tag{13}$$

When considering medium city environments

$$G_r = \left[42.57 + 13.7log10(f)\right]\left[log10(hr) - 0.585\right] \tag{14}$$

For large cities,

$$G_r = 0.759hr - 1.862 \tag{15}$$

where $f$ is frequency expressed in $GHz$ $d$ is the distance between the transmitter and receiver in $Km$, $hb$ is the transmitter antenna height in meters and $hr$ is the receiver antenna height in meters.

### 3.3.5 Stanford University Interim (SUI Model)

Path loss for this model is given in (16) as presented in [22]

$$PL_{SUI} = A + 10g\log(\frac{d}{d_o}) + s \quad \text{for } f < 2GHz \tag{16}$$

$8.2dB < s < 10.6dB$

$d$ is the distance between the transmitter and receiver

$d_o$=50m $f$ is the frequency in MHz

$$A = 20\log(\frac{4\rho d_o}{l}) \tag{17}$$

$$g = a - bh_b + \frac{c}{h_b} \tag{18}$$

Where;

$h_b$ is the base station antenna height $10m < h_b < 80m$

$l$ is the wavelength expressed in meters. a, b and c are terrain factors specified in Table 2

Table 2 Terrain Parameters

| Parameter | Category A | Category B | Category C |
|-----------|-----------|-----------|-----------|
| *a* | 4.6 | 4 | 3.6 |
| *b* | 0.0075 | 0.0065 | 0.005 |
| *c* | 12.6 | 17.1 | 20 |

### 3.3.6 Ericson Model

The equation specifying path loss for this model as presented by J. Milanovic et al [26] is shown in equation (19).

$$Pl_{Ericson} = a0 + a1\log(d) + a2\log(h_b)$$
$$+ a3\log(h_b)\log(d) - 3.2(\log(11.75h_m))^2 + g(f) \tag{19}$$

$$g(f) = 44.49\log(f) - 4.78(\log(f))^2 a + b + c \tag{20}$$

The parameters $a0, a1, a2,$ and $a3$, given in equation (19) are constants, that can be tuned to best fit specified propagation conditions. The default values of $a0, a1, a2,$ and $a3$ for different environment categories are specified in Table 3

Table 3 Default values of $a0, a1, a2$ and $a3$

| Category of Area | $a0$ | $a1$ | $a2$ | $a3$ |
|------------------|------|------|------|------|
| Urban | 36.2 | 30.2 | 12.0 | 0.1 |
| Suburban | 43.20 | 68.93 | 12.0 | 0.1 |

### 3.4 Path Loss Exponent

The path loss exponent which shows the lossy nature of a particular propagation environment was computed from the measurement data for each of the areas considered. [23] presents an approach to finding the path loss exponent as shown in (21)

$$n = \frac{\sum_{i=1}^{k}(Pl_{do} - P_i) * 10\log\left(\dfrac{d}{d_o}\right)}{\sum_{i=1}^{k}\left(10\log\left(\dfrac{d}{d_o}\right)\right)^2} \tag{21}$$

Where $P_i$ is the received power at the reference distance $d_o$, $Pl_{do}$ is the path loss at the reference distance and $n$ is the path loss exponent.

## 3.5 Root Mean Square Error(RMSE)

The RMSE  which measures the difference between the signal power predicted by a model and the actual measured signal was implemented in MATLAB. It served as a measure of accuracy to compare forecasting errors of the different propagation models given the drive test measurement data. It is defined mathematically by equation (22)

$$RMSE = \sqrt{\frac{\sum_{k=1}^{k}\left(p_i - \overset{\Lambda}{p_i}\right)^2}{k}}$$
(22)

Where $p_i$ represents the measured power value at a specified distance, $\overset{\Lambda}{p_i}$ is the predicted power value at a specified distance, k represents the number of measured samples.

## 4. RESULTS

The results presented are two-fold. The first is at an operating frequency of 800MHz and the second at an operating frequency of 2600MHz.Results are  Validated at the end  of this section

## 4.1 Results at 800MHz

The average received power was computed for each of the measurement environments by averaging the readings taken at the three different sector antennas of the base stations. The mean received power for the different environments was compared and analyzed by plots against varying distances from the base station using MATLAB. This is shown in figure 2.

As can be observed from the graph, the Received power decreases as distance away from the Base station is increased. Deviations from this trend, however, occurred on a few occasions. This was partly as a result of obstacles and a  contribution from the terrain of those environments.

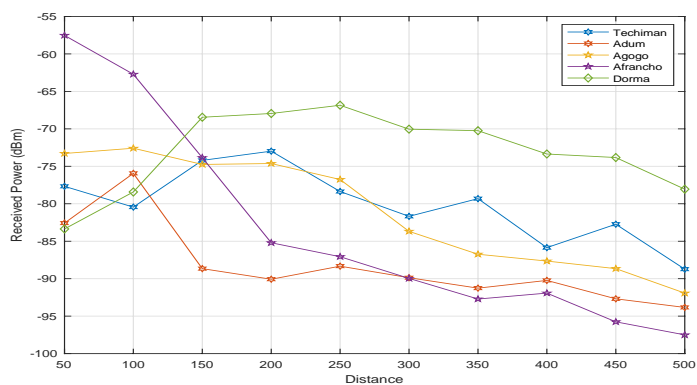

Figure 2 Received power of all Sites

### 4.1.1    Path Loss of Measured data

The experienced path loss at each measurement location at a distance d(m) was computed as follows;

Pl(dB) = EIRP(dBm)-Pr(dBm)
(23)

where $Pr$ = Mean received power in dBm, EIRP = Effective isotropic radiated power in dBm.

EIRP is given in (24) and (25)

EIRP = Pt + G-L                               (24)

Where G stands for Gains and L for losses

Typical gains considered are the antenna gains both at the transmitter and receiver end
Typical losses are connector, body and combiner loss
Expanding this yields,
EIRP = Pt + Gt + Gr-Lco-Lcon-Lbo             (25)

Pt = Transmit power (dBm),Gt = Gain of Transmit Antenna (dBi),Gr = Gain of Receive antenna (dBi),Lcon = Connector loss (dB),Lbo = Body loss (dB),Lco = Combiner loss(dB).

The Values of the stated parameters commonly applied in LTE Networks are given by S. A. Mawjoud [24]  as;

$P_{bts}$ = 40W = 46dBm   , $G_{bts}$ = 18.15dBi, $G_{ms}$ = 0dBi ,Lbo = 3dB,

 Lcon = 4.7dB, Lco = 3dB

These parameters are substituted into equation (25)

EIRP = 53.5dBm

The path loss is obtained by substituting the calculated value of EIRP (dBm) and the mean received power Pr (dBm) into equation (23).

The effect of varying distance on Path loss for each measurement environment was investigated by plots of path loss versus distance and the graph shown below in figure 3 illustrates this.

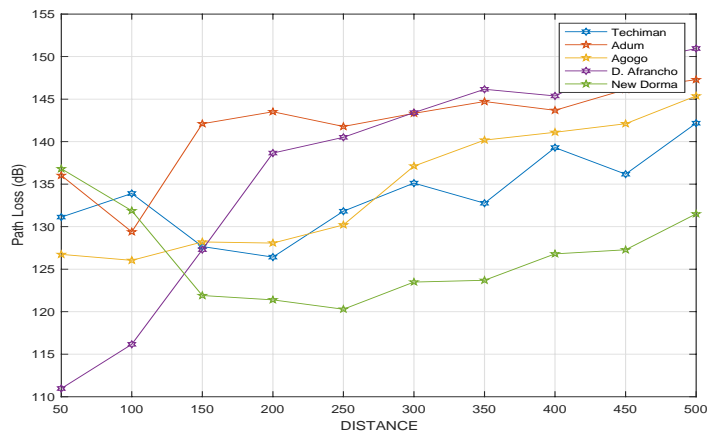

Figure 3 Path loss of all environments

It can be observed from the graph given in figure 3 that Path loss increases as the distance from the Base station increases. Comparing the path loss experienced for all the measurement environments, the Path loss of Adum and Afrancho is relatively higher compared to the other areas. The hilly nature of Afrancho and the presence of many high-rise buildings in Adum are good reasons to support the high path loss in these areas.

### 4.1.2 Comparison of Path Loss Measurement Results with Propagation Models

The path loss of each measurement environment was compared with the path loss estimations of the understudied propagation models at 800MHz for both urban(Adum and Techiman) and suburban scenarios (Agogo, Afrancho & Dorma).

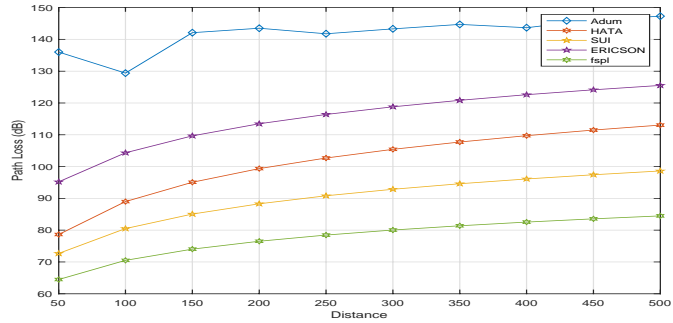

Figure 4 Path Loss of Adum Compared With Path Loss of Propagation Models

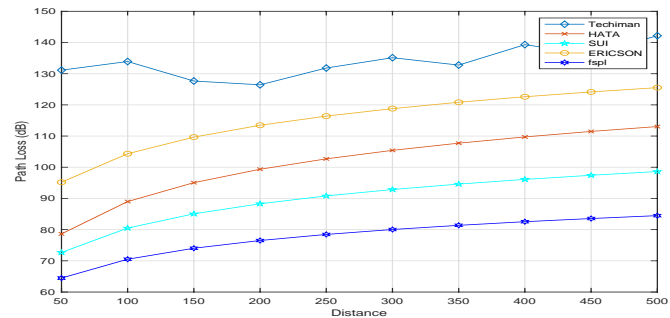

Figure 5 Path loss of Techiman compared with path loss of propagation models

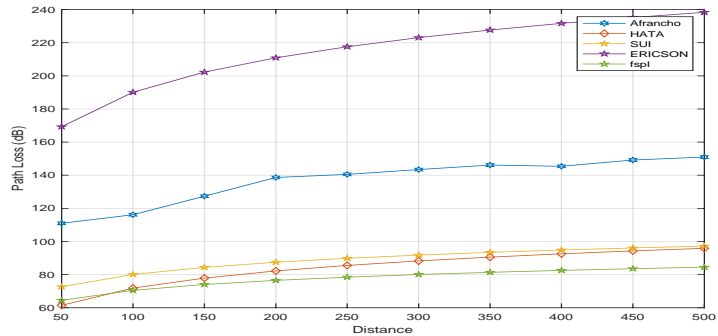

Figure 6 Path loss of Afrancho compared with path loss of propagation models

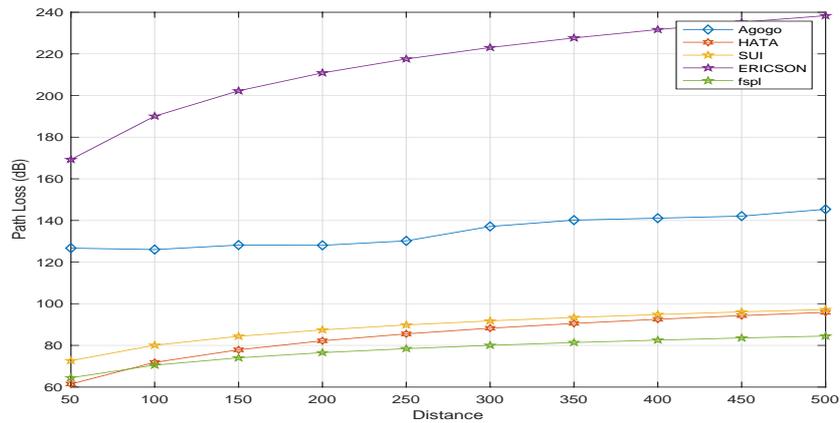

Figure 7 Path loss of Agogo compared with path loss of propagation models

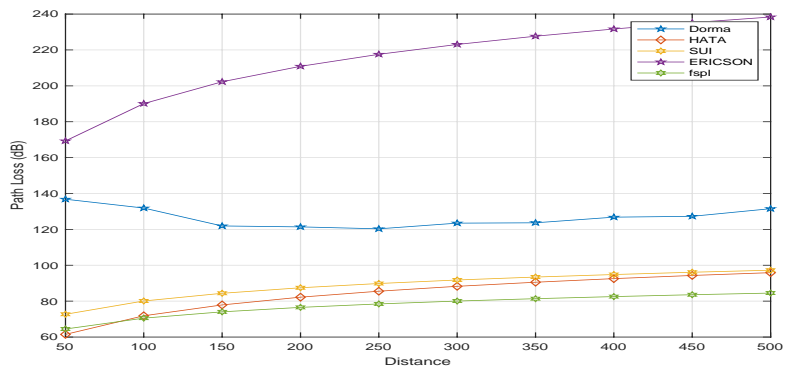

Figure 8 Path loss of New Dorma compared with path loss of propagation models

### 4.1.3 Choice of Propagation Model that best fits Measurement data

Root Mean square error was used as a quantitative measure of accuracy for choosing the propagation model that best fits the measured data in the Ghanaian environment. The best propagation model was the model that had the least Root Mean squared errors (Least RMSE). The RMSE computed for the measurement areas together with the various propagation models are given in Table 4.

Table 4 RMSE Values

| Root mean square error(urban) | | | |
|---|---|---|---|
| Environments | Hata model | SUI model | Ericson model |
| Techiman | 30.66 | 32.96 44.41 | 17.98 |
| Adum | 40.64 | 40.39  52.31 | 25.17 |
| Root mean square error(suburban) | | | |
| Agogo | 50.22 | 45.88 | 173.99 |
| Afrancho | 52.39 | 48.48 | 171.18 |
| New Dorma | 44.03 | 39.21 | 182.86 |

The Ericson model had the lowest RMSE values in Urban environments as shown in Table 4. This model is therefore chosen as the model that predicts best in Urban areas in Ghana and it is further modified and improved for more accurate predictions. In the suburban environments, the SUI model had the lowest RMSE values and hence was chosen as the best model for path loss prediction in suburban cities in Ghana. It is also further modified for a more accurate prediction.

### 4.1.4  Modification of Ericson Model

The Ericson model which best fit measurement in the urban environments was chosen and modified to fit the measured data in urban environments. To modify and further improve the Ericson model the mean square error between the urban environments and the Ericson model was added to the standardized Ericson path loss equation.

$$a0 + a1*\log(d) + a2*\log(h_b) + a3*\log(h_b)*\log(d) -3.2*(\log(11.75*h_r))^2 + g_f + RMSE \qquad (26)$$

RMSE for Adum =17.98

Adding the *RMSE* yields;

$$a0 + a1*\log(d) + a2*\log(h_b) + a3*\log(h_b)*\log(d) -3.2*(\log(11.75*h_r))^2 + g_f + 17.98 \qquad (27)$$

This new equation with the RMSE added was plotted with measurement data from Adum together with the initial standardized Ericson model equation and the graph is shown in figure 9. It can be observed that adding the *RMSE* to the initial equation improves the accuracy of prediction as the modified Ericson equation fits best with the measured data.

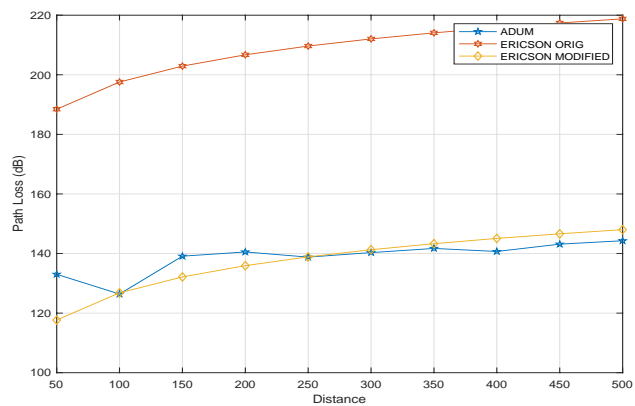

Figure 9 Comparison of the modified model and original Ericson model for Adum

The values of various parameters $a0, a1, a2, a3, h_b$, $h_r$ and $g_f$ in the Ericson model suited for an urban area were substituted into the modified equation and approximated to make the Ericson equation simple and less tedious to use yet not compromising accuracy. The resulting equation is as in equation (28)

$$Pl_{modified} = 68.30 + 30.2\log(d) + 0.139\log(d) \qquad (28)$$

A similar analysis was carried out for Techiman and figure 10 shows the modified Ericson model fitting closely to measurement data from Techiman

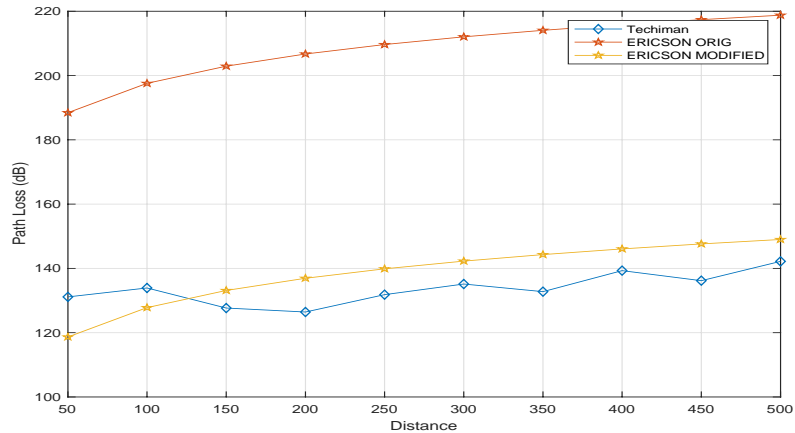

Figure 10 Comparison of the modified model and original Ericson model for Techiman

#### 4.1.5    Modification of SUI Model

**R**esults of RMSE for suburban areas favored the SUI model which had the lowest RMSE values. On this basis, the SUI model was chosen as the best-fit propagation model for path loss estimation in suburban areas in Ghana. It was further modified for more accurate predictions in Ghanaian suburban environments. The RMSE each of Agogo, Afrancho and New Dorma were added to the original SUI equations  and further simplified in equations (30) - (32)

$$PL_{SUI} = A + 10\gamma \log\left(\frac{d}{d_0}\right) + s \pm RMSE \qquad (29)$$

1) modified model for Agogo

$$Pl_{sui(simple)} = 72.68 + 24.54 \log\left(\frac{d}{d_0}\right) + RMSE \qquad (30)$$

2) modified model for Afrancho

$$Pl_{sui(simple)} = 72.68 + 24.54 \log\left(\frac{d}{d_0}\right) + RMSE \qquad (31)$$

3) modified model for New Dorma

$$Pl_{sui(simple)} = 72.6830 + 47.54 \log\left(\frac{d}{d_0}\right) + RMSE \qquad (32)$$

A graph comparing the performance of the modified model for Agogo with the original SUI model is shown in figure 11

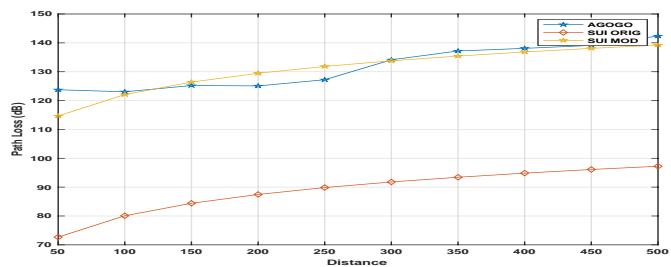

Figure 11 Comparison of modified SUI model and original SUI model for Agogo

### 4.2 Results at 2600MHz

Comparing the path loss predictions of propagation models at 2600MHz with drive test measurements of five different propagation environments, the ECC-33 model predicted close to the measurement data. This model was further modified to predict more accurately the path loss in these environments. This is shown in figures 12 & 13

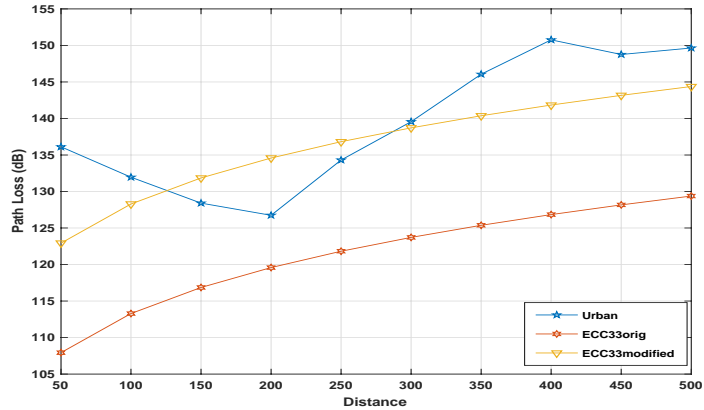

Figure 12 Comparison of modified ECC-33 model and original ECC-33 model for urban environments

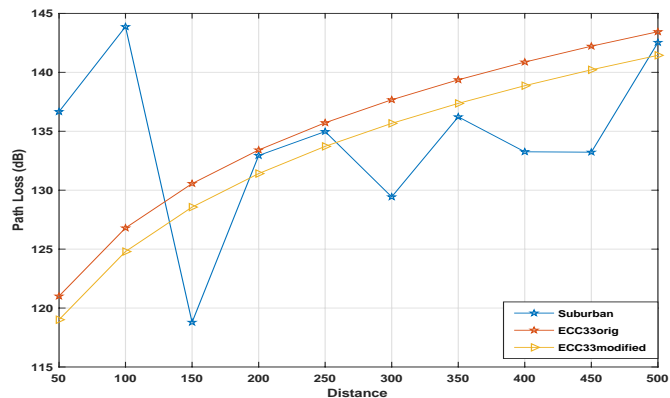

Figure 13 Comparison of modified ECC-33 model and original ECC-33 model for suburban environments

### 4.3 VALIDATION

The developed models in this study were validated by calculating the error between the measured and estimated path loss for the various measurement environments using the modified equations presented. This is achieved by using equation (28) programmed in MATLAB. The values of RMSE closer to zero indicate a better fit [25],[11]. Thus, the developed models are described as valid and suitable for the tested environments since the RMSE between the measured and the predicted path loss values are closer to zero than the initial RMSE values. Tables 7 & 8 show the RMSE generated using the developed models in this thesis at 800MHz and 2600MHz respectively.

Table 7 RMSE Values using developed models at 800MHz

| Root Mean Square Error Values of Developed Models at 800MHz | |
|---|---|
| **Measurement Environment** | **Root Mean Square Error** |
| Adum | 11.8494 |
| Techiman | 9.6717 |
| Agogo | **5.2510** |
| Afrancho | 7.1129 |
| New Dorma | 29.8491 |

Table 8 RMSE values using developed models at 2600MHz

| Root Mean Square Error Values of Developed Models at 2600MHz | |
|---|---|
| **Measurement Environment** | **Root Mean Square Error** |
| Site 1 | 9.1408 |
| Site 2 | 13.3313 |
| Site 3 | 16.8445 |
| Site 4 | 15.2780 |
| Site 5 | 11.9498 |

The improved models developed were further compared with the path loss simulated by the use of the NYUSIM simulator. This is a simulation tool developed by the New York University (NYU) wireless team and relies on huge amounts of true measured data at mm-wave frequencies in New York[27]. The simulator incorporates the CI propagation model [28]. Developed models show consistent prediction behavior compared with the NYU simulator's path loss and hence can be considered valid models for use in the Ghanaian environment with similar environmental features as the measurement environments.

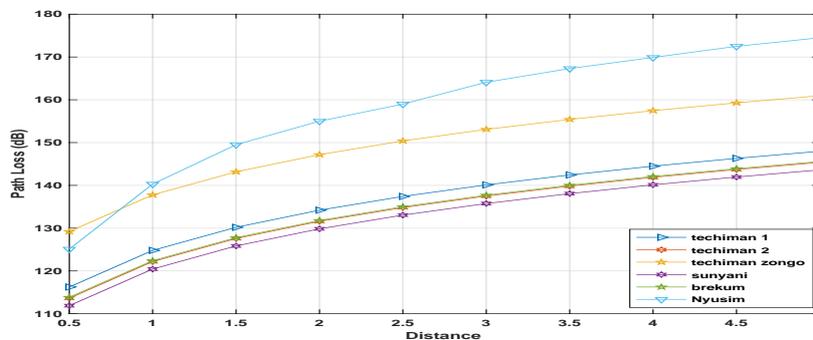

Figure 14 Comparison of Performance of Developed Models against NYUSIM at 2600MHz

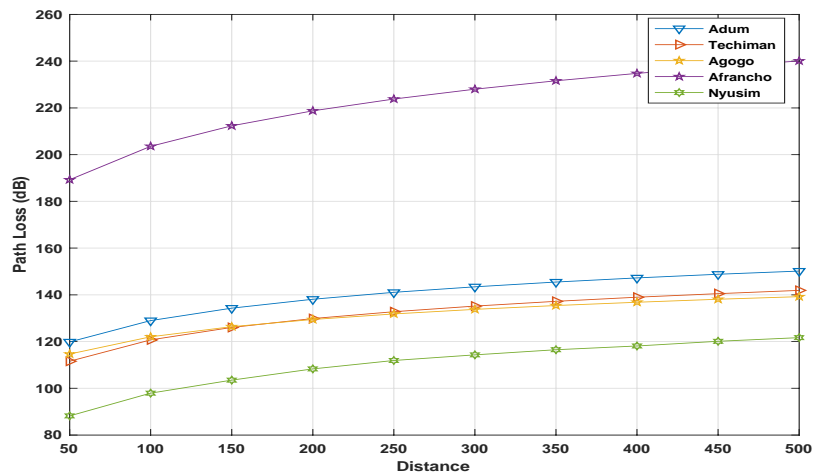

Figure 15 Comparison of Performance of Developed Models against NYUSIM at 800MHz

# 5. CONCLUSION

This study was focused on developing improved versions of industry-standard propagation models suited for LTE path loss prediction in the Ghanaian environment. Path loss of four propagation models was compared with Path loss of propagation measurements taken from five LTE 800MHz base stations located in the urban and suburban areas of Ghana using MATLAB. Results confirmed the initial assumption of the study, that propagation models predict far from the ideal. The Ericson model showed satisfactory performance in the urban environments at 800MHz. This model however over predicted the path loss in the suburban environments. The SUI model outperformed the other models in predicting close to the propagation measurement in suburban areas at 800MHz.The Ericson and SUI models were further improved for a more accurate prediction of LTE path loss in urban and suburban Ghanaian environments at 800MHz.

For similar studies at an LTE frequency of 2600MHz, the Path loss of five propagation models was compared with the Path loss of propagation measurements taken at five base stations located in the urban and suburban areas of Ghana. The ECC-33 model best fit propagation measurements both in the urban and suburban environments and hence it was developed further for use in LTE path loss estimation at 2600MHz

From the results presented, measurement data ascertained the fact that propagation models predict far from the ideal. The modified equations presented in this paper can be used in Ghanaian settings having similar characteristics with the Areas considered in this paper by network operators for accurate and simplified path loss prediction.

# REFERENCES

[1] Cisco, V. N. I. (2018). Cisco visual networking index: Forecast and trends, 2017–2022. White Paper, 1.

[2] National Communication Authority Accessed date:8th July 2018. https://www.nca.org.gh/ industry-data-2/market-share-statistics-2/data-3/

[3] World Population Review. (2019). Ghana population. Retrieved 10th May 2019 fromhttp://worldpopulationreview.com/countries/ghana-population

[4]  Nimako, S. G., & Mensah, A. F. (2012). Motivation for customer complaining and non-complaining behavior towards mobile telecommunication services. Asian Journal of Business Management, 4(3), 310-320.

[5]  Rappaport, T. S. (1996). Wireless communications: principles and practice (Vol. 2). New Jersey: prentice hall PTR.

[6]  Chebil, J., Lawas, A. K., & Islam, M. D. (2013). Comparison between measured and predicted path loss for mobile communication in Malaysia. World Applied Sciences Journal, 21, 123-128.

[7]  Mollel, M. S., & Kisangiri, M. (2014). Comparison of empirical propagation path loss models for mobile communication. Computer Engineering and Intelligent Systems, 5(9), 1-10.

[8]  Ibhaze, A. E., Imoize, A. L., Ajose, S. O., John, S. N., Ndujiuba, C. U., & Idachaba, F. E. (2017). An empirical propagation model for path loss prediction at 2100MHz in a dense urban environment. Indian Journal of Science and Technology, 10(5), 1-9.

[9]  Ajose, S. O., & Imoize, A. L. (2013). Propagation measurements and modeling at 1800   MHz in Lagos Nigeria. International Journal of Wireless and Mobile Computing, 6(2), 165-174.

[10] Imoize, A. L., Ibhaze, A. E., Nwosu, P. O., & Ajose, S. O. (2019). Determination of Best-fit Propagation Models for Pathloss Prediction of a 4G LTE Network in Suburban and Urban Areas of Lagos, Nigeria. West Indian Journal of Engineering, 41(2).

[11] Halifa, A., Tchao, E. T., & Kponyo, J. J. (2017). Investigating the Best Radio Propagation Model for 4G-WiMAX Networks Deployment in 2530MHz Band in Sub-Saharan Africa. arXiv preprint arXiv:1711.08065.

[12] Abhayawardhana, V. S., Wassell, I. J., Crosby, D., Sellars, M. P., & Brown, M. G. (2005, May). Comparison of empirical propagation path loss models for fixed wireless access systems. In 2005 IEEE 61st Vehicular Technology Conference(Vol. 1, pp. 73-77).

[13] Roslee, M. B., & Kwan, K. F. (2010). Optimization of Hata propagation prediction model in suburban area in Malaysia. Progress in Electromagnetics Research, 13, 91-106.

[14] Kale, S., & Jadhav, A. N. (2013). An Empirically Based Path Loss Models for LTE Advanced Network and Modeling for 4G Wireless Systems at 2.4 GHz, 2.6 GHz, and 3.5 GHz. International Journal of Application or Innovation in Engineering & Management (IJAIEM), 2(9), 252-257.

[15] Ekpenyong, M., Isabona, J., & Ekong, E. (2010). On Propagation Path Loss Models For 3-G Based Wireless Networks: A Comparative Analysis. Computer Science & Telecommunications, 25(2).

[16] Faruk, N., Popoola, S. I., Surajudeen-Bakinde, N. T., Oloyede, A. A., Abdulkarim, A., Olawoyin, L. A., ... & Atayero, A. A. (2019). Path Loss Predictions in the VHF and UHF Bands Within Urban Environments: Experimental Investigation of Empirical, Heuristics and Geospatial Models. IEEE Access, 7, 77293-77307.

[17] Khan, I., Eng, T. C., & Kamboh, S. A. (2012). Performance Analysis of Various Path Loss Models for Wireless Network in Different Environments. International Journal of Engineering and Advanced Technology (IJEAT) ISSN, 2249-8958.

[18] Tchao, E. T., Gadze, J. D., & Agyapong, J. O. (2018). Performance evaluation of a deployed 4G LTE network. arXiv preprint arXiv:1804.05771.

[19] Unterhuber, P., Pfletschinger, S., Sand, S., Soliman, M., Jost, T., Arriola, A., ... & Rodríguez, C. (2016). A survey of channel measurements and models for current and future railway communication systems. Mobile Information Systems, 2016.

[20] Hata, M. (1980). Empirical formula for propagation loss in land mobile radio services. IEEE transactions on Vehicular Technology, 29(3), 317-325.

[21] Neskovic, A., Neskovic, N., & Paunovic, G. (2000). Modern approaches in modeling of mobile radio systems propagation environment. IEEE Communications Surveys & Tutorials, 3(3), 2-12.

[22] Sulyman, A. I., Nassar, A. T., Samimi, M. K., MacCartney, G. R., Rappaport, T. S., & Alsanie, A. (2014). Radio propagation path loss models for 5G cellular networks in the 28 GHz and 38 GHz millimeter-wave bands. IEEE Communications Magazine, 52(9), 78-86.

[23] Zhou, T., Sharif, H., Hempel, M., Mahasukhon, P., Wang, W., & Ma, T. (2009, October). A deterministic approach to evaluate path loss exponents in large-scale outdoor 802.11 WLANs. In 2009 IEEE 34th Conference on Local Computer Networks (pp. 348-351).

[24] Mawjoud, S. A. (2013). Path loss propagation model prediction for GSM network planning. International Journal of Computer Applications, 84(7).

[25] Wu, J., & Yuan, D. (1998, September). Propagation measurements and modeling in Jinan city. In Ninth IEEE International Symposium on Personal, Indoor and Mobile Radio Communications (Cat. No. 98TH8361) (Vol. 3, pp. 1157-1160).

[26] Milanovic, J., Rimac-Drlje, S., & Bejuk, K. (2007, December). Comparison of propagation models accuracy for WiMAX on 3.5 GHz. In 2007 14th IEEE International Conference on Electronics, Circuits, and Systems (pp. 111-114).

[27] Rappaport, T. S., Sun, S., & Shafi, M. (2017, September). Investigation and comparison of 3GPP and NYUSIM channel models for 5G wireless communications. In 2017 IEEE 86th Vehicular Technology Conference (VTC-Fall) (pp. 1-5).

[28] Sun, S., MacCartney, G. R., & Rappaport, T. S. (2017, May). A novel millimeter-wave channel simulator and applications for 5G wireless communications. In 2017 IEEE International Conference on Communications (ICC) (pp. 1-7).